\documentclass[12pt,english,aps,tightenlinesletterpaper,superscriptaddress,secnumarabic,nofootinbib]{revtex4}
\usepackage[T1]{fontenc}
\usepackage[latin9]{inputenc}
\usepackage[letterpaper]{geometry}
\usepackage{amsmath}
\usepackage{graphicx}
\usepackage{amssymb}
\usepackage{esint}

\makeatletter

\providecommand{\tabularnewline}{\\}

\usepackage{babel}

\makeatother

\usepackage{babel}

\makeatother

\usepackage{babel}

\begin{document}

\title{Gravity Waves as a Probe of Hubble Expansion Rate During An Electroweak Scale Phase Transition}

\author{Daniel J.~H.~Chung}
\email{danielchung@wisc.edu}
\author{Peng Zhou}
\email{pzhou@wisc.edu}
\affiliation{{\small Department of Physics, University of Wisconsin-Madison} \\
 {\small 1150 University Avenue, Madison, WI 53706, USA}}
\begin{abstract}
Just as big bang nucleosynthesis allows us to probe the expansion rate
when the temperature of the universe was around 1 MeV, the measurement
of gravity waves from electroweak scale first order phase transitions
may allow us to probe the expansion rate when the temperature of the
universe was at the electroweak scale. We compute the simple
transformation rule for the gravity wave spectrum under the scaling
transformation of the Hubble expansion rate. We then apply this
directly to the scenario of quintessence kination domination and show
how gravity wave spectra would shift relative to LISA and BBO
projected sensitivities.
\end{abstract}
\maketitle

\section{Introduction}

The detection of gravity waves (GWs) generated during a scalar
sector's first order phase transition (PT) represents an interesting
future possibility
\cite{Steinhardt:1981ct,Witten:1984rs,Kosowsky:1991ua,Kosowsky:1992rz,Kamionkowski:1993fg,Apreda:2001us,Kosowsky:2001xp,Dolgov:2002ra,Nicolis:2003tg,Caprini:2006jb,Grojean:2006bp,Randall:2006py,Huber:2007vva,Caprini:2007xq,Megevand:2008mg,Espinosa:2008kw,Huber:2008hg,Caprini:2009fx,Caprini:2009yp,Kusenko:2009cv,Megevand:2009ut,Ashoorioon:2009nf,Das:2009ue,Kahniashvili:2009mf,Kehayias:2009tn,Durrer:2010xc}.
During a first order PT, bubbles of true vacuum nucleate, stir up the
cosmological fluid, and collide, producing GWs.  First order PTs at
the electroweak scale in particular have received much attention
because of their possible connection to a well motivated electroweak
baryogenesis scenario in which the baryon asymmetry is generated
during an electroweak phase transition (EWPT)
\cite{Shaposhnikov:1987tw,Kuzmin:1985mm}.  Studies of this scenario
are particularly timely given that Tevatron and the LHC are actively
probing the Higgs sector responsible for electroweak symmetry
breaking. If the Higgs boson or any degree of freedom that can be
responsible for a first order PT at the electroweak scale is found in
the ongoing experiments, future experiments may eventually be able to
measure all of the parameters necessary to give an accurate prediction
for the GWs.  It is important to emphasize that even if the
electroweak symmetry breaking is not a first order PT, a typical
beyond the standard model scalar sector has multiple degrees of
freedom, and some of these can undergo a first order PT at a
temperature near the electroweak scale.

Just as the relative isotope abundance measurements have led to a constraint
on the expansion rate during big bang nucleosynthesis, the measurement
of GWs may allow us to constrain any non-standard expansion
rate during the time of EWPT. Several detailed
computations of the gravity wave spectrum exist, and each give varying
degrees of dependence on the Hubble expansion rate. However, to our
knowledge, previous work does not sufficiently discuss the general dependence
of the Hubble expansion rate to directly answer the following question:
how would the observed gravity wave spectrum change if the Hubble
expansion during the EWPT was changed from
that of pure relativistic degrees of freedom?

We compute a simple transformation rule for the gravitational wave
spectrum in terms of $\xi\equiv H_{*}/H_{U}$, where $H_{U}$ is the
expansion rate which assumes radiation domination and $H_{*}$ is the
actual expansion rate:\begin{equation} \frac{d\rho_{GW}(k)}{d\ln
    k}\rightarrow\frac{1}{\xi^{2}}\frac{d\rho_{GW}(k/\xi)}{d\ln
    k}.\label{eq:main-result}\end{equation} where
$\frac{d\rho_{GW}(k)}{d\ln k}$ is the spectrum computed assuming
radiation domination.  This immediately implies the following: 1) The
peak frequency of the spectrum will shift from the standard scenario
frequency $f_{p}$ as $f_{p}\rightarrow\xi f_{p}$, and 2) the peak
amplitude of the spectrum will be suppressed from the standard
scenario amplitude $\mathcal{A}_{p}$ as
$\mathcal{A}_{p}\rightarrow\mathcal{A}_{p}/\xi^{2}$.
The intuition for the amplitude is that less source
contributes to the gravitational wave at a typical spacetime point
today because of the smaller intersection of the past null boundary
with the approximately compact time support of the source. The
intuition for the frequency shift is that all conformal symmetry
breaking scales relevant for the observable frequency range is
controlled by the Hubble expansion rate.

We then apply this scaling relationship to the results of
\cite{Huber:2008hg,Caprini:2007xq} to compute how the gravity wave
spectrum will transform due to the assumption of the existence of a
quintessence kination dominated phase
\cite{Caldwell:1997ii,Peebles:1987ek,Kamionkowski:1990ni,Salati:2002md,Rosati:2003yw,Profumo:2003hq,Pallis:2005hm}.
Such assumptions are interesting because as pointed out by
\cite{Salati:2002md} (related scenarios were also suggested before by
\cite{Kamionkowski:1990ni,Barrow:1982ei}), the freeze-out abundance of
thermal relics can be strongly enhanced in scenarios in which the
energy density is dominated by the kinetic energy of the quintessence
field (kination domination) during the time of freeze-out, but dilutes
away by the time of big bang nucleosynthesis (BBN). Such kination
dominated freeze-out scenarios are then consistent with standard
cosmology and predict that the standard relic abundance computed from
the parameters extracted from collider measurements will be mismatched
from the relic abundance deduced by observational cosmology. Thus,
this scenario has interesting implications for physics models with
thermal dark matter candidates (e.g. models with low energy
supersymmetry such as the minimal supersymmetric extension of the SM
(MSSM), technicolor models, models with large/warped extra dimensions,
or certain classes of little Higgs models), which will be probed at
the LHC and other experiments in the foreseeable future (for collider
implications of this class of models, see for example
\cite{Chung:2007cn}).  Furthermore, as pointed out by
\cite{Chung:2007vz,Profumo:2004ty}, large annihilation cross sections
compatible with the dark matter explanation of the excess positrons
\cite{Adriani:2008zr,Beatty:2004cy} can be compatible with the right
thermal relic abundance since the effective boost factor coming from
the kination scenario can easily be as large as $10^{3}$.  Finally,
since the measurement of CMB B-mode polarization can almost model
independently falsify this scenario, this scenario can be nontrivially
checked with a variety of probes.  In particular, the EWPT gravity
wave probe in conjunction with dark matter cosmology can represent a
smoking gun probe of the scenario if the gravity wave is measurable
\cite{Chung:2007vz} and colliders can eventually measure the requisite
short-distance parameters with sufficient accuracy.

The order of presentation is as follows.  In the next section, we
present the main analytic result of this paper.  Sec.~\ref{sec:survey}
focuses on checking explicit consistency of our result with the
existing literature on explicit gravity wave spectrum computations.
In Sec.~\ref{sec:kinationdominationexample}, we apply the
transformation relation to the quintessential kination scenario and
give plots showing how the transformed spectra look relative to the
projected sensitivities of LISA and BBO.  Sec~\ref{sec:caveat}
discusses all of the caveats associated with the analytic result.  We
then conclude with a summary.  The appendices contain some of the
details used throughout the paper.

As far as conventions are concerned, we use the reduced Planck's constant
$M_{p}=2.4\times10^{18}$ GeV and also assume a flat FRW background
metric $ds^{2}=a^{2}(t)(dt^{2}-|d\vec{x}|^{2}).$

\section{\label{sec:generalanalyticarg}General Analytic Arguments}

In this section, we compute how the gravitational wave spectrum will
transform under the situation that during the last part of the PT,
the Hubble expansion rate is $\xi H_{U}$ where $H_{U}$ is what the
expansion rate would be in the {}``usual'' radiation domination
epoch. Because the main arguments rely only on the general form of
the gravity wave equation and dimensional analysis, the transformation
results will be very robust and nearly model independent. In a latter
section, we check the consistency of the transformation rules with
explicit model dependent computations in the literature.

Consider the transverse-traceless perturbation $h_{ij}^{TT}$ about the
background metric:\begin{equation}
  ds^{2}=a^{2}(t)\left[dt^{2}-(\delta_{ij}+h_{ij}^{TT})dx^{i}dx^{j}\right].\end{equation}
Using the pseudotensor expression for the energy density in gravity
waves, the energy density in GWs can be expressed as\begin{equation}
\rho_{GW}=\frac{M_{p}^{2}}{4a^{2}}\langle\partial_{0}h_{ij}^{TT}\partial_{0}h_{ij}^{TT}\rangle.\end{equation}
During a phase transition (PT) the energy density in the gravity wave
can be written as\begin{align} \rho_{GW}(\vec{x},t)= &
\frac{1}{M_{p}^{2}}\left(\frac{a_{*}}{a}\right)^{4} \langle\left\{
\frac{\partial}{\partial t}[\Lambda_{ij,lm}\int
  d^{4}x'G_{ret}(x;x')T_{lm}(x')]\right\}
^{2}\rangle|_{PT}\label{eq:rhogwexact}\\ \sim &
\frac{1}{M_{p}^{2}}\left(\frac{a_{*}}{a}\right)^{4}\left\langle
\frac{\partial}{\partial
  t}\left(\frac{1}{\square}T_{ij}\right)\frac{\partial}{\partial
  t}\left(\frac{1}{\square}T_{ij}\right)\right\rangle
|_{PT}\label{eq:rhoGW}\end{align} where $a_*$ is the scale factor at
the beginning of the PT, $G_{\mbox{ret}}(x,t;x',t')$ is the Minkowski
retarded Green's function \begin{equation} \square
  G_{ret}(x;x')=\left[\frac{\partial^{2}}{\partial
      t^{2}}-\nabla^{2}\right]G_{ret}(x;x')=\delta^{4}(x-x')\end{equation}
and the transverse traceless non-local projection operator is defined
as \begin{align} \Lambda_{ij,lm} &
  =P_{il}P_{mj}-\frac{1}{2}P_{ij}P_{lm}\\ P_{ij} &
  =\delta_{ij}-\frac{1}{\nabla^{2}}\partial_{i}\partial_{j}\end{align}
where $T_{ij}$ is the conformal coordinate stress tensor and the symbol
$|_{PT}$ represents the evaluation at the end of the phase
transition. Evaluation of the right hand side is accomplished through
the power spectrum $P(k_{1},t_{1}',t_{2}')$ written as\begin{equation}
\left\langle
\tilde{T}_{ij}(t_{1}',\vec{k}_{1})\tilde{T}_{ij}^{*}(t_{2}',\vec{k}_{2})
\right\rangle
=(2\pi)^{3}\delta^{(3)}(\vec{k}_{1}-\vec{k}_{2})P(k_{1},t_{1}',t_{2}')\left[\rho_{f}^{\mbox{rest}}\gamma_{v_f}^{2}v_{f}^{2}\right]^{2}
a_*^2\end{equation} which incorporates spatial translational
invariance and a suggestive normalization of bubble pressure squared
that assumes that the correlator is dominated by disconnected diagrams
(i.e.  bubble interactions are neglected).  Here,
$\rho_{f}^{\mbox{rest}}$ is a fiducial energy density of the fluid
measured by an observer at rest with the fluid at the location of the
bubble wall, $\gamma_{v_f}$ is the usual Lorentz contraction factor
associated with the fluid velocity $v_{f}$ behind the bubble wall. It
is related to the bubble wall velocity $v_{w}$ by
$v_{f}=(v_{w}-c_{s})/(1-v_{w}c_{s})$, where $c_{s}$ is the speed of
sound and $v_{w}$ is taken to be a constant \cite{Moore:2000wx}. We
therefore find the gravity wave spectrum as\begin{equation}
\frac{d\rho_{GW}}{d\ln k}
=\frac{1}{(2\pi)^{2}}\frac{1}{M_{p}^{2}}\left(\frac{a_{*}}{a}\right)^{4}\left[\rho_{f}^{\mbox{rest}}\gamma_{v_f}^{2}v_{f}^{2}\right]^{2}
a_*^2 \int
dt_{1}'dt_{2}'\cos\left[k(t_{1}'-t_{2}')\right]\left[k^{3}P(k,t_{1}',t_{2}')\right]
\end{equation}
Note that the integration is assumed to be over all time, but
$k^{3}P(k,t_{1}',t_{2}')$ will have support dominantly over a time
period $\Delta t$ surrounding the time of the PT.  As we will check
explicitly later, this turns out to be reasonable even for relatively
longer lived turbulent sources.  The characteristic sizes governing
$k^{3}P$ include the duration of the last part of the PT $a_* \Delta
t<\frac{1}{H_{*}}$ (where $H_*$ is the Hubble expansion rate at the
time of the phase transition), the time $t_{*}$ at which the PT
occurs, and the size of the typical bubble $R$. Since\begin{equation}
\frac{1}{R}\sim\frac{1}{v_{w}a_*\Delta
  t}\label{eq:sizeofbubble}\end{equation} if we assume that these are
the only important scales in the spectrum and that $t_{*}$ only enters
as \begin{equation}
  t_{i}'-t_{*},\label{eq:dependenceontstar}\end{equation} we can
write\begin{equation} F_{k\Delta t}((t_{1}'-t_{*})/\Delta
t,(t_{2}'-t_{*})/\Delta t)\equiv
k^{3}P(k,t_{1}',t_{2}').\label{eq:powerspec-scaling}\end{equation}
Caveats to this assumption will be discussed in Sec.~\ref{sec:caveat},
but the general conclusions there will be that this assumption is
robust.  This gives us\begin{eqnarray} \frac{d\rho_{GW}}{d\ln k} & = &
\frac{a_*^2}{(2\pi)^{2}M_p^2}\left(\frac{a_{*}}{a}\right)^{4}\left[\rho_{f}^{\mbox{rest}}\gamma_{v_f}^{2}v_{f}^{2}\right]^{2}\int
dt_{1}'dt_{2}'\cos\left[k(t_{1}'-t_{2}')\right]F_{k\Delta
  t}(\frac{t_{1}'-t_{*}}{\Delta t},\frac{t_{2}'-t_{*}}{\Delta t}) \\ &
= &
\frac{\left(a_* \Delta t\right)^2}{(2\pi)^2M_{p}^{2}}\left(\frac{a_{*}}{a}\right)^{4}\left[\rho_{f}^{\mbox{rest}}\gamma_{v_f}^{2}v_{f}^{2}\right]^{2}\int dq_{1}'dq_{2}'\cos\left[k\Delta
  t(q_{1}'-q_{2}')\right]F_{k\Delta t}(q_{1}',q_{2}')\end{eqnarray}
where $q_{i}'$ is integrated over $(-\infty,\infty)$ and is
dimensionless.  Hence, the only conformal symmetry breaking scale in
the integrand is $\Delta t$.

To proceed with the analysis, we need to determine what sets the mass
dimension scale of $\Delta t$. As is well known \cite{Turner:1992tz},
the nucleation rate of the PT bubble per
unit volume per unit time at temperature $T$ is \begin{equation}
\gamma=C_{1}T^{4}\exp\left[\left(-S_{*}^{(3)}-(t-t_{*})\frac{dS^{(3)}}{dt}|_{t_{*}}\right)/T\right]\label{eq:nucleationrate}\end{equation}
where $S^{(3)}$ corresponds to the appropriate bounce action at finite
temperature and $C_{1}$ is assumed to be $\mathcal{O}(1)$. Hence,
\begin{eqnarray}
\frac{dS^{(3)}}{dt}|_{t_{*}} & = & \frac{\dot{T}}{T}\frac{dS^{(3)}}{d\ln T}|_{t_{*}}\nonumber \\
 & = & -\left[\frac{a(t_*) H_{*}}{1+\frac{1}{3}\frac{d\ln g_{*S}}{d\ln T}}\right]\frac{dS^{(3)}}{d\ln T}|_{t_{*}}\label{eq:wherehubbleenters}\end{eqnarray}
where $H_{*}$ is the expansion rate at the
time $t_*$ of the PT and we have assumed that the entropy
whose density proportional to $g_{*S}T^{3}a^{3}$ is conserved where
$a$ is the scale factor and $g_{*S}$ counts the entropy degrees
of freedom. As given by Eq.~(\ref{eq:durationofPT}) of Appendix
\ref{sec:Bubbles-Filling-Space}, the completion of the PT
occurs during a time interval of \begin{equation}
a_* \Delta t\propto\frac{1}{H_{*}}
\label{eq:proportionality}
\end{equation}
for $H_* a_* \Delta t < 1$.  Hence, the only Hubble expansion rate dependence in the integrand
\begin{equation}
\cos\left[k\Delta t(q_{1}'-q_{2}')\right]F_{k\Delta t}(q_{1}',q_{2}')\end{equation}
 is in $k\Delta t$.

Explicitly, with the definition\begin{equation}
\xi\equiv\frac{H_{*}}{H_U},\end{equation}
 where the $U$ subscript stands for {}``usual'' radiation domination
scenario, the integral \begin{equation}
\int dq_{1}'dq_{2}'\cos\left[k\Delta t(q_{1}'-q_{2}')\right]F_{k\Delta t}(q_{1}',q_{2}')\end{equation}
 is invariant under the transformation\begin{equation}
\Delta t\rightarrow\Delta t/\xi\end{equation}
 \begin{equation}
k\rightarrow k\xi\end{equation} Hence, with the present assumptions,
the spectrum changes under the transformation $H_U \rightarrow H_U \xi$
as\begin{eqnarray}
\frac{1}{(2\pi)^2}\frac{1}{M_{p}^{2}}\left(\frac{a_{*}}{a}\right)^{4}\left[\rho_{f}^{\mbox{rest}}\gamma_{v_f}^{2}v_{f}^{2}\right]^{2}\left(a_*
\Delta t\right)^{2}\int dq_{1}'dq_{2}'\cos\left[k\Delta
  t(q_{1}'-q_{2}')\right]F_{k\Delta t}(q_{1}',q_{2}') &
\rightarrow\nonumber
\\ \frac{1}{(2\pi)^2}\frac{1}{M_{p}^{2}}\left(\frac{a_{*}}{a}\right)^{4}\left[\rho_{f}^{\mbox{rest}}\gamma_{v_{f}}^{2}v_{f}^{2}\right]^{2}\left(a_*
\Delta
t/\xi\right)^{2}\int dq_{1}'dq_{2}'\cos\left[k\Delta
  t(q_{1}'-q_{2}')/\xi\right]F_{k\Delta
  t/\xi}(q_{1}',q_{2}')\label{eq:spectralchange}\end{eqnarray} or
equivalently\begin{equation} \frac{d\rho_{GW}(k)}{d\ln
  k}\rightarrow\frac{1}{\xi^{2}}\frac{d\rho_{GW}(k/\xi)}{d\ln
  k}.\label{eq:spectrum shift}\end{equation} This is the main analytic
result of this paper.  Although this result in some sense has been
reported in the literature indirectly before (as we survey below), one
of the points of this study is to examine the robustness of the
relationship and to spell out the assumptions necessary.

From this equation, we can thus extract two easy to remember features: 
\begin{enumerate}
\item If the peak frequency without quintessence is $k_{p}$, the new
  peak frequency is at $k_{p}\xi$.  This is intuitive from recognizing
  that a smaller bubble length scale results if the expansion is
  faster (corresponding to $\xi>1$) since the bubble length scale has
  a smaller time to grow before the PT is completed.
\item The amplitude of the spectrum at the peak should decrease as $1/\xi^{2}$
compared to case with $\xi=1$. The intuition for this result is that
less source contributes to the gravitational wave at a fixed point
because of the past Minkowski null boundary and the approximately
compact time support of the source. 
\end{enumerate}
Eq.~(\ref{eq:spectrum shift}) is essentially a classical dimensional
analysis coming from the assumptions summarized by
Eqs.~(\ref{eq:sizeofbubble}) and (\ref{eq:dependenceontstar}), and the
result is useful in allowing us to read off how the gravity wave
spectrum depends on the general assumptions of the Hubble expansion
during the PT.  As we shall see in the literature
survey, the details of the spectrum for frequencies larger than the
peak frequency is difficult to compute and very uncertain.
Nonetheless, we will show in sections \ref{sec:survey} and
\ref{sec:caveat} that our scaling arguments above are robust, and
Eq.~(\ref{eq:spectrum shift}) will most likely apply to improved
spectra that will be derived by all future more accurate computational
techniques.

Note that the amplitude dependence on the fluid energy
$\rho_{f}^{\mbox{rest}}$ written in Eq.~(\ref{eq:spectralchange}) in a
very intuitive form is a bit misleading since it naively looks as one
can increase the amplitude of the gravity wave by increasing this
quantity.  However, since $\rho_{f}^{\mbox{rest}}$ scales as the
radiation energy density, when the gravity wave energy density is
compared with the radiation energy density, one power of it is
normalized away.  The second power of $\rho_{f}^{\mbox{rest}}$
actually represents the clock units with which to measure $(a_* \Delta t)^2$
since when divided by $M_p^2$, it represents the approximate expansion
rate squared of a radiation dominated universe.  

Hence, it is not really the absolute magnitude of
$\rho_{f}^{\mbox{rest}}$ that is important for increasing the
measurable gravity wave amplitude but the dimensionless quantity $(a_*
\Delta t)^2 \rho_{f}^{\mbox{rest}}/M_p^2$.  As we will see explicitly
in Sec.~\ref{sec:kinationdominationexample}, obtaining a spectrum
observable at the LISA experiment will require the duration $a_*
\Delta t$ of the PT to be of the same order of magnitude as the
expansion rate determined by the radiation energy density during the
PT.  Physically, this effectively corresponds to a sufficiently large
potential barrier suppressing bubble nucleation such that a
non-negligible supercooling occurs before the PT completes.  Although somewhat tangential to the point of our paper, we
discuss this issue a bit further in
Sec.~\ref{sec:kinationdominationexample} and Appendix
\ref{sec:TdepofS3}.
   
Although our next goal is to utilize Eq.~(\ref{eq:spectrum shift}) to
make predictions for the quintessence kination dominated scenario, we
will in the next section first check the consistency of our result
with some of the explicit computations in the literature.  There, we
will also consider turbulence contributions to the gravity wave
production and show that even when the turbulent source is long lived,
its dominant contribution will be from the phase transition time
period, allowing Eq.~(\ref{eq:spectrum shift}) to remain a good
approximation.  If the reader is not interested in the consistency
check, the reader is encouraged to skip to
Sec.~\ref{sec:kinationdominationexample}.

\section{\label{sec:survey}Survey Of Explicit Computations and Detailed Analysis}

In this section, we will survey the literature which computes the
gravity wave spectrum using both simulations and analytic techniques
\cite{Kosowsky:1991ua,Kosowsky:1992rz,Kamionkowski:1993fg,Caprini:2007xq,Caprini:2009fx,Caprini:2006jb,Caprini:2009yp,Huber:2007vva,Huber:2008hg,Huber:2000mg}.
Our aim is to show that our scaling assumptions resulting in
Eq.~(\ref{eq:spectrum shift}) are consistent with the existing
explicit computations.

References \cite{Caprini:2007xq,  Caprini:2009fx, Huber:2007vva, Huber:2008hg}
find the following generic result regarding GWs generated
from bubble collision. The spectrum is found to have a rising and
a falling shape, where the increasing side of the spectrum scales
as $k^{3}$ and the peak position is at a wave vector of order of
$1/R_{*}$ ($R_{*}$ is the typical bubble size at the end of the
PT). Physically, one can attribute the $k^{3}$ scaling
law to the compactness of the sources and the spatial homogeneity
of their distribution. At the same time, one can understand the appearance
of the conformal symmetry breaking scale $1/R_{*}$ as being the only
identifiable classical length scale in the problem. While there is no well
known uncertainty about the rising part of the spectrum \cite{Caprini:2009fx},
there is a large uncertainty associated with the falling part (UV
part) of the spectrum. Direct simulations find that it scales as $k^{-1}$\cite{Huber:2008hg},
while the analytic calculations give model-dependent results \cite{Caprini:2007xq,Caprini:2009fx}. 

In addition to bubble collisions, turbulent motion of the fluid during the
EWPT will also generate GWs \cite{Caprini:2006jb,  Caprini:2009yp, Kosowsky:2001xp, Dolgov:2002ra, Nicolis:2003tg}.
One can estimate the turbulence spectrum with either dimensional analysis
or the velocity correlation function \cite{LL}. It is possible that
fluid turbulence and magnetohydrodynamic (MHD) turbulence can
contribute to GW production as much as bubble collisions. 

We now consider the details below.

\subsection{Simulation for bubble collision}

GW generation from bubble collisions in zero temperature vacuum has been
extensively studied by Kosowsky et al.~\cite{Kosowsky:1991ua}. Each
bubble is represented as an $O(3,1)$ kink solution to the scalar field
($\phi$) equation of motion. The latent energy released in the phase
transition is turned into the bubble wall's gradient energy
$(\nabla\phi)^{2}$ and kinetic energy $(\partial_{t}\phi)^{2}$. Using
numerical simulations, it is found that the GW spectrum depends
primarily on the large scale features of the source. Specifically, a
model of the source's stress energy tensor which ignores the collision
region between the two bubbles still results in a GW spectrum that is
almost identical to the one from a full simulation. This ``envelope
approximation'' when applied to many bubbles collisions
\cite{Kosowsky:1992rz} ($N\sim20-200$) results in a GW spectrum
characterized by a peak frequency and a ratio of GW total energy over
the released latent energy as 
\begin{equation}
  \omega_{max}=1.6\beta,\quad
  E_{GW}/E_{vac}=0.06(H_{*}/\beta)^{2}
\label{eq:vacuumbubblescaling}
\end{equation} 
where $\beta$ is a parameter controlling the bubble nucleation rate
$\gamma$ as in $\gamma\propto\exp(\beta a_* t)$, and $\beta$ is related to
the PT duration by $\beta\sim \left( a_* \Delta t \right)^{-1}$.
Upon going through their arguments, one can show that $H_*$ (Hubble
expansion rate at the time of the PT) in this equation
can be seen to be a reparameterization of radiation temperature $T_*$
rather than the expansion rate of the universe itself.\footnote{Even
  though their bubble computations are based on zero temperature
  vacuum, they add in the expansion of the universe by hand to scale
  the energy density appropriately for the physical values today.
  This is the source of the $T_*$ dependence.}
Hence, Eq.~(\ref{eq:vacuumbubblescaling}) is consistent with our
result that the fraction of the energy of the GW is proportional to
the PT duration $(a_* \Delta t)^{2}$.
A statistical approach is also considered in \cite{Kosowsky:1992rz},
where the gravity wave spectrum is computed as an incoherent sum over
individual bubbles weighted by the bubble size distribution
function. In the many bubble case, the high frequency part of the
spectrum is enhanced by the multiplicity of small bubbles.

Building on the zero temperature work, the finite temperature
situation is studied in \cite{Kamionkowski:1993fg}. Finite temperature
modifies the Higgs boson's effective potential and introduces a fluid
dynamical degree of freedom. If the PT produces large
latent heat, the bubble wall velocity would be supersonic, leading to
a process called detonation. As the detonation front expands, the
fluid that is swept by the front is being compressed and dragged along
(see \cite{Steinhardt:1981ct} for details of computing the single
bubble's velocity profile and temperature profile). A full simulation
of two bubble collision is computationally difficult, due to the
chaotic fluid motion and the wide range of length scales involved. To
circumvent this difficulty, {}``envelope approximation'' is again
applied, resulting in the following total GW energy fraction and peak
frequency \cite{Kamionkowski:1993fg}:

\begin{align}
\Omega_{GW}h^{2} &
\approx1.1\times10^{-6}\kappa^{2}\left(\frac{H_{*}}{\beta}\right)^{2}\left(\frac{\alpha}{1+\alpha}\right)^{2}\left(\frac{v_{w}^{3}}{0.24+v_{w}^{3}}\right)\left(\frac{100}{g_{*}}\right)^{1/3}\label{eq:KamionkowskiEq}\\ f_{max}
&
\approx5.2\times10^{-8}\mbox{Hz}\left(\frac{\beta}{H_{*}}\right)\left(\frac{T_{*}}{1\mbox{
    GeV}}\right)\left(\frac{g_{*}}{100}\right)^{1/6}\end{align} where
$v_{w}$ is the detonation front's velocity, $T_{*}$ is the phase
transition temperature, and $g_{*}$ is the relativistic degree of
freedom during the EWPT.  Here $\alpha=\rho_{vac}/\rho_{rad}$ is the
ratio of vacuum energy density to the radiation energy density, which
characterizes the strength of the PT.  The variable
$\kappa\sim\rho_{fluid}/\rho_{vac}$ is an efficiency factor
quantifying the fraction of the available vacuum energy that goes into
the kinetic motion of the fluid.  Evidently, the GW spectrum's
amplitude and peak position have the same scaling dependence on
$\beta$ as the zero temperature case, which is consistent with our
scaling result since $\beta$ quantifies $1/(a_* \Delta t) \propto H$
while $H_*^2$ here represents the radiation energy density and not the
total expansion rate.  (Even though $H_*$ here corresponds to $H_U$ in
other sections of the paper, we maintain the original literature's
notation in this  survey section to emphasize the
non-transparency of the scaling relationship in the literature.)
Hence, if exotic fluid component contributions to the stress tensor
increase the expansion rate during the PT keeping the temperature
fixed, then the scaling can be read off from the existing literature
by scaling with $\beta$.  In that sense, the change in the peak
position and amplitude resulting from our Eq.~(\ref{eq:spectrum
  shift}) is not particularly new.  On the other hand, the robustness
of this simple scaling relationship for the entire observable spectrum
and its application to the kination dominated quintessence scenario
have not been explored before this paper, to our knowledge.
Generalizing the two bubble collision simulation to many bubbles
($N\sim100)$ \cite{Huber:2008hg}, the GW spectrum's high frequency
part is found to be enhanced from $k^{-1.8}$ to $k^{-1}$, again due to
the small bubble effect at the end of phase transition.

\subsection{Analytic calculation for Bubble Collision}

Reference \cite{Caprini:2007xq} uses the stochasticity of the source
to estimate the gravity wave's spectrum. Two assumptions are made
about the velocity field in the bubble collision:
\begin{enumerate}
\item The velocity field's distribution is approximately Gaussian, i.e.
the four points correlator $\langle vvvv\rangle$ is determined by
the two point correlator$\langle vv\rangle$.
\item The two point correlator $\langle v(x)v(y)\rangle$ is nonzero if
$x$ and $y$ can belong to the same bubble. 
\end{enumerate}
The following spectrum is thereby obtained: 
\begin{equation}
\frac{d\Omega(k,\eta_{0})h^{2}}{d\ln
  k}\approx\frac{3}{2\pi^{3}}\left(\frac{g_{0}}{g_{*}}\right)^{\frac{1}{3}}\Omega_{\mbox{rad}}h^{2}\left(\frac{\Omega_{\mbox{kin}}^{*}}{\Omega_{\mbox{rad}}^{*}}\right)^{2}\left(\frac{H_{*}}{\beta}\right)^{2}\frac{(1-s^{3})^{2}}{s^{4}}\times\frac{0.21\left(\frac{Z}{Z_{m}}\right)^{3}}{1+\left(\frac{Z}{Z_{m}}\right)^{2}+\left(\frac{Z}{Z_{m}}\right)^{4.8}}\label{eq:CDS-spectrum}
\end{equation}
where $s$ is the ratio of bubble wall thickness to radius and $Z/Z_m \sim
k/k_p$ is a dimensionless wave-number. \footnote{We use $k$ and
  $\omega$ interchangeably in the GW spectrum.}  Since the amplitude
contains the factor
$\left(\frac{H_{*}}{\beta}\right)^{2}$, and the
frequency dependence is through a function of $Z=kL_{*}$ (where $L_*$
is the typical length scale at the end of the stirring phase), the
spectrum's parametric scaling is consistent with our scaling rule
result in Eq.\,(\ref{eq:spectrum shift}).

\subsection{Scaling of the Turbulence Generated Gravity Wave Spectrum}

As the thermal bubbles percolate, the fluid within the bubbles
collides and generates turbulence. Even though a detailed simulation
of turbulence's evolution is difficult, some statistical features can
be derived from dimensional analysis and intuition. Generally, the
turbulence contains eddies of different sizes, as the larger ones
break down to smaller ones, energy is also cascaded down to smaller
scales. In a fully developed turbulence, the standard intuitive
assumptions are that the energy cascade rate is a constant over time
and is also a constant for different scales
\cite{LL,Kamionkowski:1993fg}. For example, if the cascade rate
$\epsilon$ were not constant on different scales, there cannot be a
steady state since energy would be building up at a particular length
scale. However, as far as eddy sizes are concerned, since a largest
scale $L$ fixed by the bubble size and a smallest scale fixed by the
viscosity exist, the energy cascade rate cannot be exactly scale
invariant. Assuming that a turbulent eddy of size $l$ with velocity
$v_{l}$ breaks down in a few turn over times $\tau_{l}\sim l/v_{l}$
(which is true by dimensional analysis assuming that viscosity plays a
negligible role), the energy cascade rate per unit mass for
non-relativistic eddies is \begin{equation}
  \epsilon\sim\frac{v_{l}^{2}}{l/v_{l}}\sim v_{l}^{3}/l,\mbox{ or
    equivalently }v_{l}\sim(\epsilon l)^{1/3}.\end{equation} It is
interesting that the assumptions of $\epsilon$ being a constant over
different length scales and viscosity being unimportant until very
short length scales fix the velocity spectrum by dimensional analysis.
As $l$ decreases (i.e. for smaller eddies), the dissipation effect
becomes increasingly important, and the energy dissipation rate per unit
mass is given by \begin{equation} \nu(\nabla v)^{2}
  \sim\nu\left(\frac{v_{l}}{l}\right)^{2}\end{equation} where $\nu$ is
the kinematic viscosity. If $l$ is decreased to the point that the
energy dissipation rate equals the energy cascade rate, the turbulence
would cease to exist. This scale is the Kolmogorov microscale
$\lambda$: \begin{equation}
  \epsilon\sim\nu\left(\frac{v_{\lambda}}{\lambda}\right)^{2}\Longrightarrow\lambda\sim
  L(Lv_{L}/\nu)^{-3/4}\sim L(Re)^{-3/4}\end{equation} where the
dimensionless Reynolds number $R_{e}\equiv Lv_{L}/\nu$ relate the
largest scale $L$ and the smallest scale $\lambda$. In the case of
EWPT, the Reynolds number is usually on order of $10^{13}$, causing
$\lambda$ to be negligible compared to $L$, indeed.

Given these two parameters $L$ and $v_{L}$, the energy spectrum
of a fully developed turbulence can be estimated as \begin{equation}
E(k)\sim\epsilon^{2/3}k^{-5/3},\mbox{ for }\lambda<k^{-1}<L\end{equation}
which is used in \cite{Kamionkowski:1993fg} to estimate a GW spectrum
\begin{equation}
\frac{\omega}{\rho}\frac{d\rho_{GW}}{d\omega}\approx\left(\frac{H_*}{\beta}\right)^{2}vv_{0}^{6}\left(\frac{\omega}{\omega_{0}}\right)^{-9/2}.\end{equation}
In the formula, the parameter $v_{0}$ is the typical fluid velocity at
the length scale of the largest bubble size $L$, not to be confused
with the bubble boundary velocity $v$, and
$\omega_{0}\sim\tau_{L}^{-1}\sim\beta v^{-1}v_{0}$.  This spectrum is
valid up to the smallest scale of turbulence, i.e.
$k^{-1}\sim\lambda$. The factors $\left(\beta\right)^{-2}$ and
$\omega/\omega_{0}$ in the spectrum indicate that the turbulence
generated gravity wave spectrum also is consistent with our scaling
rule equation Eq.~(\ref{eq:spectrum shift}).  Just as for all the
previous examples, $H_*$ should be viewed as parameterizing the
critical temperature $T_*$ and not the true expansion rate of the
scale factor.

Reference \cite{Caprini:2009yp} also considers the initial stirring
phase and the final decay phase of the turbulence. The velocity
correlator function of the turbulence is used to find the
stress-energy correlation functions. It is found that both the MHD and
fluid turbulence last for a long time after the PT has
ended, in contrast with the bubble collision case. The gravity wave
spectrum is found to be\begin{align}
\frac{d\Omega_{GW}h_{0}^{2}}{d\log k} &
=12(2\pi)^{2}C_{s}^{2}\Omega_{rad,0}h_{0}^{2}\left(\frac{g_{0}}{g_{fin}}\right)^{1/3}\left(\frac{\Omega_{S*}}{\Omega_{rad*}}\right)K_{*}^{3}\nonumber
\\ &
\times\left[\int_{0}^{1}dy\frac{y^{3\gamma+2}}{y+\frac{t_{in}}{\tau_{L}}}I_{s}(K_{*},y,y)\int_{y}^{y_{top}}\frac{dz}{z+\frac{t_{in}}{\tau_{L}}}\cos\left(\frac{\pi
    K_{*}}{v_{L}}(z-y)\right)\right.\label{eq:CDSturb}\\ &
  +\left.\int_{1}^{y_{fin}}dy\frac{y^{-7\gamma}}{y+\frac{t_{in}}{\tau_{L}}}I_{s}(K_{*},y,y)\int_{y}^{y_{top}}\frac{dz}{z+\frac{t_{in}}{\tau_{L}}}\cos\left(\frac{\pi
    K_{*}}{v_{L}}(z-y)\right)\right].\nonumber \end{align} In this
spectrum, the subscript 0 denotes the present time, $*$ denotes the
end of PT, and $fin$ denotes the end of the
turbulence. The subscript $s$ in $C_{s}$ can be either $v$ or $b$,
which stands for the fluid turbulence or magnetic turbulence. $C_{s}$
is a numerical factor and $\Omega_{s}$ is the corresponding source's
energy fraction of the total energy density. The wave-vector $k$ is
rendered dimensionless as $K_{*}=kL_{*}$.  The variable $y$ is a time
variable normalized by the largest eddy turn over time $\tau_{L}$:
$y=\frac{t-t_{in}}{\tau_{L}}$ where $t_{in}$ is the beginning time of
the stirring phase.  The finish time of the turbulence is denoted as
$y_{fin}(k)$, the k-dependence indicates that different modes end at
different times.  The integral of $z$ 's upper limit
$y_{top}\equiv\min[y_{fin},y+\frac{v_{L}x_{c}}{\pi K_{*}}]$ serves as
a cut-off of correlation between sources at different
time. $I_{s}(K_{*},y,y)$ is the normalized dimensionless equal-time
correlator of the source.  The index $\gamma$ is explained below.

The first line represents an overall normalization, controlled by
the amplitude of the source. The second line with time integral $\int_{0}^{1}dy...$
represents the contribution to GW from turbulence during the stirring
up time period, i.e. the PT period. The last line with
time integral $\int_{1}^{y_{fin}}dy...$ represents the contribution
to GW during the decay of the turbulence. We shall show the evolution
of the turbulence's stress energy tensor only depends on one scale,
i.e.\,the PT duration $\Delta t$, which is also the
largest eddy turn over time $\tau_{L}$. In the stirring up part,
$\tau_{L}$ is clearly the only scale. In the free decay part, there
might be a new time scale controlling the decay, but we shall see
there is none. The decay of the turbulence is modeled as\begin{equation}
\frac{\text{\ensuremath{\Omega}}_{T}}{\Omega_{rad}}\sim(\frac{t-t_{in}}{\tau_{L}})^{-5\gamma}.\end{equation}
This power law relation with time is a scale free relation. Therefore,
both parts of the evolution of the turbulence contain at most one
scale. On the other hand the correlation between sources at different
times is also modeled in a scale free way. It is assumed that the
source with a certain wave vector $k$ at two different times $t_{1}$and
$t_{2}$ are uncorrelated if the time lapse is larger than a few oscillation
time, i.e. \begin{equation}
\langle\tilde{T}(k,t_{1})\tilde{T}(k,t_{2})\rangle=0,\,\,\mbox{if
}|t_{1}-t_{2}|\gtrsim \frac{x_{c}}{k}\,\,,x_{c}\sim O(1)\end{equation}
Thus the evolution and correlation of the source contains no other
scale than $\tau_{L}$. Finally, one can schematically put the above
Eq.~(\ref{eq:CDSturb}) into the following form: \begin{align*}
 & \rho_{src}\times\int_{-\infty}^{+\infty}dt_{1}\int_{-\infty}^{+\infty}dt_{2}P(kL_{*},t_{1},t_{2})\\
\propto & (\Delta t)^{2}\int_{-\infty}^{+\infty}\frac{dt_{1}}{\Delta t}\int_{-\infty}^{+\infty}\frac{dt_{2}}{\Delta t}\tilde{P}(kL_{*},\frac{t_{1}}{\Delta t},\frac{t_{2}}{\Delta t})\end{align*}
where $\tilde{P}(kL_{*},\frac{t_{1}}{\Delta t},\frac{t_{2}}{\Delta t})$
is a scale free formula. Thus, the GW spectrum in the long lasting
turbulence case is still consistent with our assumptions leading to
our scaling rule equation
Eq.~(\ref{eq:spectrum shift}).

\section{\label{sec:kinationdominationexample}Example: Kination Domination Phase of Quintessence}

In this section we will apply Eq.~(\ref{eq:spectrum shift}) to the
results of \cite{Huber:2008hg,Caprini:2007xq} to compute how the
gravity wave spectrum will shift due to the assumption of the
existence of a quintessence kination dominated phase
\cite{Caldwell:1997ii,Peebles:1987ek,Kamionkowski:1990ni,Salati:2002md,Rosati:2003yw,Profumo:2003hq,Pallis:2005hm}.
The class of models that we are interested in can be described by the
Lagrangian\begin{equation} \mathcal{L}=\frac{1}{2}(\partial
q)^{2}-V(q)+\mathcal{L}_{M}-\frac{M_{p}^{2}}{2}R\end{equation} where
the real scalar field $q$ couples only to gravity (described by the
Ricci scalar $R$) and the matter sector $\mathcal{L}_{M}$ (which must
contain the Standard Model sector) through the minimal metric
coupling. As discussed for example in \cite{Chung:2007vz}, the
quintessence energy density scaling with $a$ is not strongly
constrained during the early universe if one is willing to tune
$V(q)$.  If the kinetic energy dominates, the phase of the $q$ field
is said to be in a kination dominated phase, and the energy density
behaves as \begin{equation}
  \rho_{q}\equiv\frac{1}{2}\dot{q}^{2}+V(q)\propto
  a^{-6}\end{equation} with an equation of state $w=1$.  Starting from
this phase, when the kinetic energy has decayed away, $P/\rho$ can
behave as $w\approx-1$ equation of state fluid during cosmological
periods for which we have empirical evidence for the existence of dark
energy. If $V(q)$ participates such that it gently pushes $q$ to
compensate for the Hubble friction, then $\rho_{q}$ can decrease less
quickly than $a^{-6}$. Instead of focusing on the details of the finely
tuned $V(q)$ that can realize different scaling behavior with $a$, we
will simply parameterize the quintessence energy decay as%
\footnote{See appendix \ref{sec:formalmap} for a formal mapping to potential.%
} \begin{equation}
\rho_{q}\propto a^{-n}\end{equation}
with $n\in\{4,5,6\}$. The results can also be used to understand situations
in which the faster expansion rate does not arise from quintessence but rather other exotic fluid components contributing to the stress-energy tensor
\cite{Barrow:1982ei}.

A large family of quintessence models can be described by two
parameters $(n,\eta)$ characterizing the energy density's scaling
behavior (defined above) with Hubble expansion and the relative relic
density at the time of BBN, respectively. The latter is defined
(e.g. \cite{Chung:2007cn,Chung:2007vz}) as \begin{equation}
  \eta\equiv\frac{\rho_{q}(t_{BBN})}{\rho_{\gamma}(t_{BBN})}.\end{equation}
Current projection of the collider sensitivity to $\eta$ if dark
matter is an MSSM LSP is $10^{-4}$ for the LHC (upper bound only) and
$10^{-6}$ for the ILC \cite{Chung:2007cn}.  As we will now see, the
gravity wave probe sensitivity is even greater.

\begin{figure}
\begin{centering}
\includegraphics[scale=0.8]{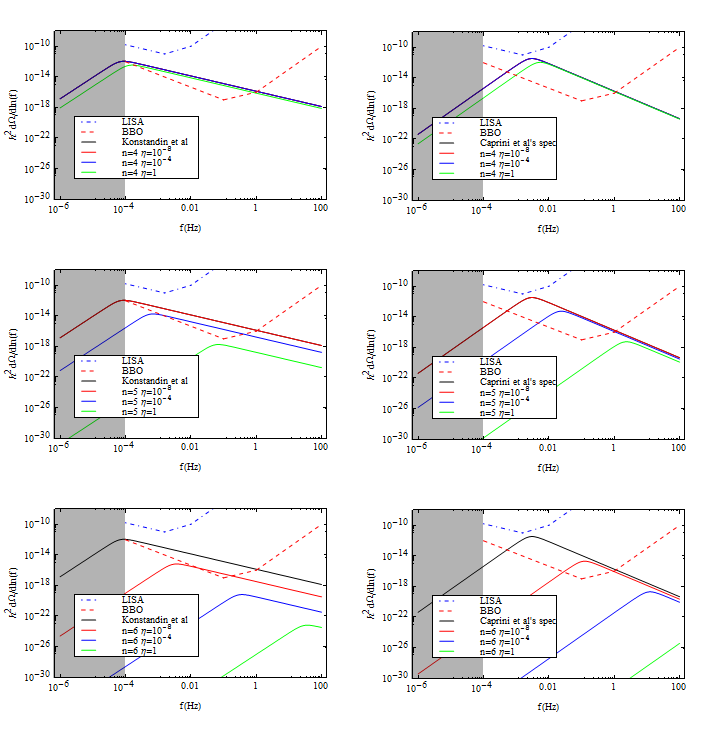}
\par\end{centering}

\caption{These plots illustrate the gravitational wave spectrum change due
to the effects of a kination phase of quintessence. The phase transition
parameters are chosen as $\alpha=0.2,T_{*}=70 \ {\rm GeV},\beta/H_{U}=30$.
On the left are plots based on the simulation of Konstandin et al.~\cite{Huber:2008hg} and on the right are plots based on that of Caprini et al.~\cite{Caprini:2007xq}. From the top row downward we have $n=4,5,6$. In each plot, we
have the detector sensitivity lines, the original spectrum line without
quintessence and the lines for $\eta=10^{-8},10^{-4},1$.\label{fig:ModGW}}
\end{figure}

We take the gravity wave spectrum from the recent literature and
compute the shifted spectrum due to quintessence. In
Fig.~\ref{fig:ModGW}, the plots in the left column are based on the
formula from numerical simulations \cite{Huber:2008hg}. Starting from
the nMSSM Higgs model with superpotential \begin{equation}
  W_{nMSSM}=\lambda\hat{S}\hat{H_{1}}\cdot\hat{H_{2}}-\frac{m_{12}^{2}}{\lambda}\hat{S}+W_{MSSM}\end{equation}
the result reported in \cite{Huber:2007vva} involved a scan over the model
parameter space looking for regions that give rise to a strong first
order PT. One set of PT parameters consists of
\{$\alpha=0.2,T_{*}=70 \mbox{ GeV}, \beta/H_*=30$\}, taken
from set (6) in Table I in \cite{Huber:2008hg}. We shall use this set
of parameters for all the plots in Fig.~\ref{fig:ModGW}.  To make the
plots on the left column of Fig.~\ref{fig:ModGW}, we
use 
\begin{equation} 
\Omega_{GW}(f) =
\tilde{\Omega}_{GW}\frac{(a+b)\tilde{f^{b}}f^{a}}{b\tilde{f}^{(a+b)}+af^{(a+b)}} 
\end{equation}
where
\begin{eqnarray}
\tilde{f} &
  =&16.5\times10^{-3}\textrm{mHz}\left(\frac{\tilde{f_{*}}}{\beta}\right)\left(\frac{\beta}{H_{*}}\right)\left(\frac{T_{*}}{100\mbox{GeV}}\right)\left(\frac{g_{*}}{100}\right)^{1/6}
  \\ 
 \frac{\tilde{f_{*}}}{\beta} &
  =&\frac{0.62}{1.8-0.1v_{b}+v_{b}^{2}} \\ 
\tilde{\Omega}_{GW} &
  = & 1.67\times10^{-5} h^{-2}\frac{0.11v_{b}^{3}}{0.42+v_{b}^{2}}\kappa^{2}\left(\frac{H_{*}}{\beta}\right)^{2}\left(\frac{\alpha}{\alpha+1}\right)^{2}\left(\frac{100}{g_{*}}\right)^{1/3}
  \\ v_{b}(\alpha) &
  =&\frac{\sqrt{1/3}+\sqrt{\alpha^{2}+2\alpha/3}}{1+\alpha}\label{eq:vw_alpha}\\ \kappa(\alpha)
  &
  =&\frac{1}{1+0.715\alpha}[0.715\alpha+\frac{4}{27}\sqrt{\frac{3\alpha}{2}}].
\label{eq:kappa_alpha}\end{eqnarray}
All the variables we do not explicitly discuss below have been defined
in an earlier part of this paper.  The quantities with tildes
correspond to the quantities evaluated at the peak of the GW spectrum,
while those with a star subscript refer to the values defined at the
time of the phase transition. The parameters $a$ and $b$ in the above
first formula correspond to the absolute values of the slopes of the
increasing and decreasing regions of the gravity wave spectrum, and
they are fit from the numerical simulations to be $a=3$ and $b=1$
\cite{Huber:2008hg}. The modified spectra are plotted by shifting the
original spectrum according to Eq.~(\ref{eq:spectrum shift}), where
$\xi$ is computed for each curve from the corresponding $(n,\eta)$
as \begin{equation}
  \xi=\sqrt{\frac{\rho_{rad}(t_{BBN})\eta\left(\frac{a(t_{BBN})}{a(t_{EWPT})}\right)^{n}+\rho_{rad}(t_{EWPT})}{\rho_{rad}(t_{EWPT})}},\mbox{
    where }\rho_{rad}(g,T)= g\frac{\pi^{2}}{30}T^{4}\end{equation}

To show that our scaling of the spectra can be applied independently
of the details of the PT computation, we plot on the
right column of Fig.~\ref{fig:ModGW} spectra based on analytic
estimation of \cite{Caprini:2007xq}: \begin{equation}
  \frac{d\Omega(k,\eta_{0})h^{2}}{d\ln
    k}\approx\frac{3}{2\pi^{3}}\left(\frac{g_{0}}{g_{*}}\right)^{\frac{1}{3}}\Omega_{\mbox{rad}}h^{2}\left(\frac{\Omega_{\mbox{kin}}^{*}}{\Omega_{\mbox{rad}}^{*}}\right)^{2}
  \left(\frac{H_{*}}{\beta}\right)^{2}\frac{(1-s^{3})^{2}}{s^{4}}\times\frac{0.21\left(\frac{Z}{Z_{m}}\right)^{3}}{1+\left(\frac{Z}{Z_{m}}\right)^{2}+\left(\frac{Z}{Z_{m}}\right)^{4.8}}\end{equation}
\begin{eqnarray}
\frac{\Omega_{\mbox{kin}}^{*}}{\Omega_{\mbox{rad}}^{*}} & =&
\frac{4}{3}\frac{(sv_{f})^{2}}{1-(sv_{f})^{2}}\\ s & =
&c_{s}/v_{b}\\ v_{f} &
=&(v_{b}-c_{s})/(1-v_{b}c_{s})\\ \Omega_{rad}h^{2} & =
&4.15\times10^{-5}\end{eqnarray} where $Z=kv_{b}/(a
\beta),Z_{m}=3.8,c_{s}=\sqrt{1/3},\; g_{0}=3.75,\; g_{*}=106.75$, and
detonation front velocity $v_{b}$ is related to $\alpha$ as usual. As
mentioned above, the PT parameters are chosen to be the
same as those for the left column plots.

The three rows correspond to the parameters $n=4,5,6$ respectively,
and the four curves within each plot from top to bottom correspond to
$\eta=0,10^{-8},10^{-4},1$ respectively.  According to the bottom row
plots, if the underlying Higgs physics is determined sufficiently
accurately at colliders, a measurement of gravity wave spectrum by the
projected BBO experiment matching the properties of the Higgs model
would rule out any kination dominated scenario explanation of
$\mathcal{O}(1\%)$ discrepancy between collider determination of
thermal relic density and cosmological measurements. More
specifically, for $\eta\ll1$, the discrepancy caused by the kination
phase can be expressed as\begin{equation}
\frac{\Delta\Omega^{(K)}}{\Omega^{(U)}}\sim10^{5}\eta\left(\frac{m_{\chi}}{100\mbox{
    GeV}}\right)^{2}\end{equation} for a dark matter particle of mass
of order 100 GeV \cite{Chung:2007cn}. As expected, it is clear from
the Fig.~\ref{fig:ModGW} that GWs have much stronger sensitivity to
$\eta$ and $n$ than collider/dark matter combination of measurements.
Since the kination scenario effectively allows a large (up to $10^3$)
boost factor reconciling enhanced galactic annihilations (such as
those relevant for PAMELA \cite{Adriani:2008zr}) with dark matter
abundance \cite{Chung:2007vz}, such measurements from BBO can strongly
constrain scenarios reconciling collider physics, cosmological DM
abundance, and indirect dark matter signals.
Even more optimistically, if one can obtain the peak position (and the
right column plot happens to give the correct picture), then one may
be able to measure both $\eta$ and $n$, which can be overconstrained
by the possible dark matter data which also is sensitive to $\eta$ and
$n$.

\begin{figure}
\begin{centering}
\includegraphics[scale=0.8]{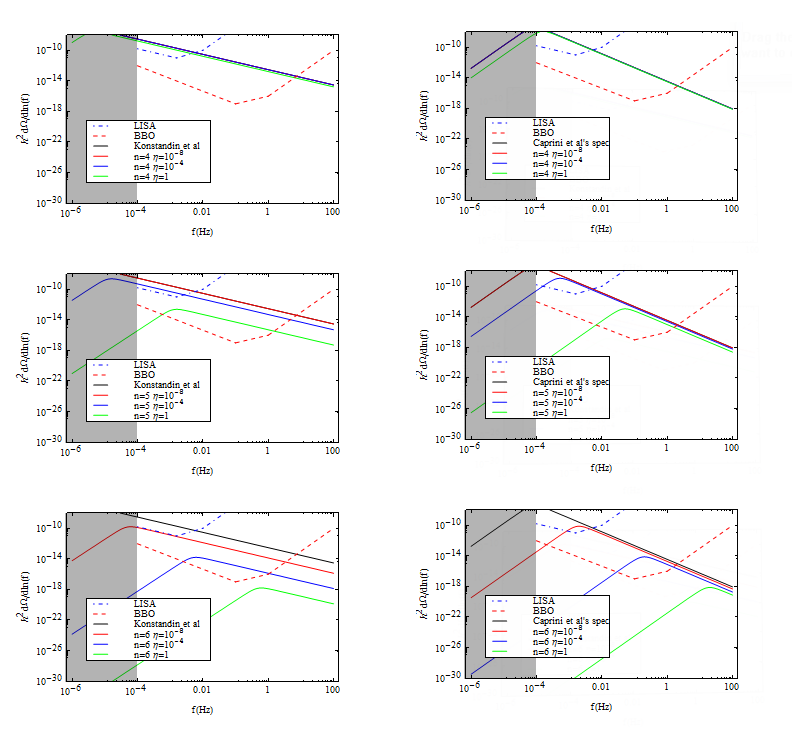}
\par\end{centering}

\caption{This figure is an analog of Fig.~\ref{fig:ModGW} except with
  the phenomenological parameters of the phase transition tuned to
  give a larger amplitude.  The parameters are
  $\beta/H_U=1$, $\alpha=1$, $T_*=70$ GeV.  Physically, this corresponds
  to a phase transition with a long duration ($\beta/H_U$ is not large)
  and a large ratio of vacuum energy density to radiation energy
  density ($\alpha=1$ instead of
  $\alpha=0.2$).  \label{fig:ModGWlongdur}}
\end{figure}

Given that the parameters used for Fig.~\ref{fig:ModGW} does not lead
to a GW spectrum that LISA can measure, one might ask ``What kind of
underlying scalar sector parameters will lead to observable
gravity waves for LISA?''  Naively, one might suspect that since
$\left(\rho_{f}^{\mbox{rest}}\right)^2$ appears in front of
Eq.~(\ref{eq:spectralchange}), one simply would have to make this
large.  However, as we discussed at the end of
Sec.~\ref{sec:generalanalyticarg}, the duration $a_* \Delta t$ of the phase
transition uses $\rho_{f}^{\mbox{rest}}$ as a clock and the two
quantities are not independent.  For example, in the absence of exotic
fluid element like quintessence, $\left( a_* \Delta t \right)^2$ scales inverse
proportionally with $\rho_{f}^{\mbox{rest}}$.  
Hence, the combination $\rho_{f}^{\mbox{rest}} a_* \Delta t$ does not
increase even if one increases $\rho_{f}^{\mbox{rest}}$ by itself.
Hence, to make the relevant combination large, one actually needs a
more difficult to achieve model building ingredient of making
$dS^{(3)}/dT$ as small as possible.  Typically, this requires small
thermal corrections to the relevant scalar field near the critical
temperature.  Keeping other parameters fixed, this corresponds to a
large $T_c$ (this is the temperature at which the PT begins and not
when most of the PT completes).  Qualitatively, this corresponds to a
situation in which the system is closer to having supercooling rather
than not.  A semi-quantitative discussion within the
context of a toy model is presented in Appendix \ref{sec:TdepofS3}.

Although we provide no underlying physics model, we plot in
Fig.~\ref{fig:ModGWlongdur} a hypothetical spectrum that would be
generated by a ``long'' PT to demonstrate the scaling effect on a
gravity wave spectrum measurable by LISA.  Because we are setting
$\beta/H=1$ for Fig.~\ref{fig:ModGWlongdur}, this is at the edge of
the validity of the scaling relationship which assumed that $a_* H_*
\Delta t < 1$.  On the other hand, the qualitative behavior of the
scaling relationship is accurate.  Note that a typical beyond the SM
Higgs sector extended by singlets go through many PTs before the EWPT,
some of which may lead to the spectrum as shown in
Fig.~\ref{fig:ModGWlongdur}.  The effect of superposing such spectra
is beyond the scope of this work.

\section{\label{sec:caveat}Caveats}

In this section, we explore the limitations of the assumptions leading
to the scaling Eq.~(\ref{eq:spectrum shift}).  The most important
assumption is that there is only one relevant dimensionful scale, the
Hubble expansion rate, during the generation of GWs. Therefore, the
existence of any other conformal symmetry breaking physics relevant
within the kinematic regime of our interest may invalidate our
argument. The possible scales include the bubble wall thickness, the
correction to the energy density due to bubble interactions, and the
dissipation scale associated with turbulence. 

The bubble wall thickness is usually defined near the phase boundary
where the energy is concentrated.  In the vacuum bubble case, the most
prevalently nucleated Higgs bubble solutions are believed to be
described by $O(3,1)$ symmetric solutions $\phi(t,r)$ which are
functions of $t^2-r^2$. In the lab frame the bubble wall's thickness
will contract towards zero as the bubble wall velocity approaches the
speed of light. Thus the bubble wall thickness is very small compared
to the bubble radius.  In the thermal bubble case, the energy is
modeled to be contained in the fluid's kinetic energy 
rather than in the Higgs field's profile.  Here the bubble wall is
defined by the portion of fluid having large enough velocity to give
an appreciable contribution to GW.  Steinhardt \cite{Steinhardt:1981ct}
showed that there is a large class of solutions in which the fluid's
speed only depends on parameter $r/t$, i.e. as the bubble expands
the velocity profile would expand accordingly. Therefore the bubble
wall thickness is proportional to the bubble size, and in particular,
it is not a scale independent of the Hubble rate.

Next we would like to consider the effect of small scale physics in
the bubble collision case. We will show below that the dimensionless
gravitational wave spectrum
\begin{equation}
k^{3}P(k,t_{1},t_{2})
\end{equation}
in the $k$ range of interest for BBO and LISA depends appreciably
 only on the dimensionless combinations 
\begin{equation} \{k\Delta
   t,\frac{t_{i}-t_{*}}{\Delta
     t},\frac{\vec{M}}{T_{*}(\vec{M})}\}\label{eq:keyparam}
\end{equation}
where $t_{*}$ is the time of the PT, $T_{*}$ is the
temperature at the time of the PT which is assumed to
depend only on the short distance physics parameters $\vec{M}$ as far
as the conformal symmetry breaking parameters are concerned, and $a_* \Delta
t\propto1/H_{*}$ is the duration of the PT proportional
to the inverse Hubble expansion rate at the time of the phase
transition. Since the last of these parameters do not scale with the
Hubble expansion rate, the conclusion of our paper is robust if
Eq.~(\ref{eq:keyparam}) are the only relevant conformal symmetry
breaking parameters.

To show this, we start with the effective classical description of the stochastic
process of bubble creation and stress tensor evolution. Suppose we
make the reasonable approximation that scalar field evolves classically
except with a stochastic source term that can create bubbles. Hence,
we write the equation of motion for the scalar sector as\begin{equation}
D_{\mu}D^{\mu}\phi_{i}+\frac{\partial V}{\partial\phi_{i}^{*}}-\frac{\partial}{\partial\phi_{i}^{*}}\mathcal{L}_{I}=J_{i}(x)\label{eq:eomwithstochasticsource}\end{equation}
where $\mathcal{L}_{I}$ is a short distance physics governed non-derivative
interaction Lagrangian, $V$ is a scalar sector potential, and \begin{equation}
J_{i}(x)=\int d^{4}y\pi(y)j_{i}(x-y)\label{eq:sourceinteg}\end{equation}
is a stochastic distribution which accounts for thermal/quantum fluctuation
induced bubble creation. The stochastic function $\pi(y)$ is the
number of bubbles created per unit time per unit volume at a small
cell volume\footnote{Bubble nucleation from a classical {}``appearance'' point of view
is a collective field process typical of soliton creation since a bubble
of radius larger than a critical radius is required to appear for
it to expand classically.%
} centered at $y$ %
 and $j_{i}(x-y)$ is a non-stochastic fixed function\footnote{Strictly speaking, it is a distribution since it may contain Dirac
delta functions.%
} representing the effective classical source for a single bubble.
Let $P(\pi)$ be the probability of obtaining a particular function
$\pi$ for a single realization of the universe. It is in principle
fixed by the path integral computation of the bubble nucleation, but
the details will not be important for the scaling arguments that we
present if the conditions we later discuss are satisfied. The ensemble
average $\left\langle T_{ij}(t_{1}',\vec{x})T_{ij}(t_{2}',\vec{y})\right\rangle $
can be written as\begin{equation}
\left\langle T_{ij}(t_{1}',\vec{x}_{1})T_{ij}(t_{2}',\vec{x}_{2})\right\rangle =\int\mathcal{D}\pi P(\pi)\left\{ T_{ij}(t_{1}',\vec{x}_{1})|_{\pi}T_{ij}(t_{2}',\vec{x}_{2})|_{\pi}\right\} .\end{equation}
where $T_{ij}$ (a functional of $\vec{\phi}$) is implicitly dependent
on $\pi$ through Eqs.~(\ref{eq:eomwithstochasticsource}) and ~(\ref{eq:sourceinteg}):\begin{equation}
\phi_{i}(x)=\phi_{i}[\pi](x).\end{equation}
Note that the functions $\pi$ that generate non-vanishing inhomogeneities
are the only important contributions, even if they may not be at
the peaks of the probability functional $P(\pi)$.

Now, we impose 3 conditions which will limit the number of conformal
breaking parameters that the system depends on:
\begin{enumerate}
\item The value of $\left\langle
  T_{ij}(t_{1}',\vec{x}_{1})T_{ij}(t_{2}',\vec{x}_{2})\right\rangle $
  needs to be known only in the interval $t_{*}<t_{i}'<t_{*}+\Delta t$
  (where $a_* \Delta t < 1/H_{*}$) to obtain an order of magnitude
  accurate gravity wave spectrum since it falls off rapidly outside of
  that time interval.  As shown in Appendix
  \ref{sec:Bubbles-Filling-Space}, the effective duration of
  the PT is \begin{equation} a_* \Delta
  t\approx\frac{1}{H}_{*}F\left(\vec{M}/T_{*}(\vec{M})\right)\label{eq:delttdefhere}\end{equation}
  where $F\ll1$ is a dimensionless function of short distance physics
  conformal symmetry breaking parameters $\vec{M}$ and the critical
  temperature $T_{*}$. To make the argument of the dimensionless
  function dimensionless, $\vec{M}$ has been scaled by the critical
  temperature $T_{*}$ which itself is assumed to be a function of
  $\vec{M}$ only and not $H_{*}$.  The dominance of the correlator
  within this short window of time is a reasonable generic assumption
  since the completion of the PT at $t_{*}+\Delta t$ corresponds to a
  homogeneous and isotropic scalar field phase which cannot source
  GWs. On the other hand, the plasma inhomogeneities can grow (albeit
  slowly because of the relativistic pressure) through gravitational
  clustering. As a first guess, this effect should be most pronounced
  on short distance physics scales (characterized by $\vec{M}$) for
  which the inhomogeneities leading to gravitational potential would
  be the largest. If this is true we can neglect these complications
  as we discuss further below. Another caveat is that $T_{*}$ is not
  necessarily independent of $H_{*}$ since a very large $H_{*}$ can
  lead to the decoupling of a particular relativistic species if that
  species has interaction rate $\Gamma<H_{*}$. In that case, the
  temperature of that species, call it $T_{1}$, can evolve differently
  from the rest of the thermal plasma. If a finite Higgs VEV can lead
  to interactions with this species at the time $t_{*}$ of the PT, one
  can have in addition to $T_{*}$ another temperature $T_{1}(t_{*})$
  which now does depend on $H_{*}$. Note that in such situations,
  $T_{*}$ and $T_{1}(t_{*})$ can be coupled and the dependence on
  $H_{*}$ enters through $T_{1}(t_{*})$. This is highly model
  dependent and as long as the number of species decoupling as a
  function $H_{*}$ are not larger than $\mathcal{O}(10)$, its effect
  will be less than order 10\%.  Leaving further investigation of
  these effects to a future work and staying consistent with state of
  the art computations in the literature, we will not discuss the
  possible breakdown of this assumption further here.
\item The function $\pi(y)$ is parametrically dependent on
  $\{\vec{M},T(t),a_* \Delta t\}$
where $T(t)$ is the time dependent mean temperature of the relativistic
fluid. Note that this is certainly true for the mean bubble nucleation
rate per unit spacetime volume $\gamma(t)$ .
\item The functional $P[\pi]$ is parametrically dependent on
  $\{\vec{M},T(t),a_* \Delta t\}$.
From the computation of the saddle point approximation of $\gamma(t)$,
this is reasonable.
\end{enumerate}
These conditions will now be used for a dimensional analysis aimed at
checking whether there are any physically relevant conformal symmetry
breaking scales that we missed.  We will in effect be reanalyzing some
of the steps in Sec.~\ref{sec:generalanalyticarg} to obtain a sense of
what kind of details can invalidate the scaling result of
Eq.~(\ref{eq:spectrum shift}).

First, let's go to the Fourier basis useful for computing the gravity
wave spectrum. Because of the FRW background assumption, the
correlators should have no preferred spatial position or
direction. Hence, we must have the form \begin{eqnarray} \left\langle
  T_{ij}(t_{1}',\vec{x})T_{ij}(t_{2}',\vec{y})\right\rangle & = &
  \left[\rho_{B}^{\mbox{rest}}\gamma_{v_{f}}^{2}v_{f}^{2}\right]^{2}a_*^2
  \int\frac{d^{3}k_{1}}{(2\pi)^{3}}e^{i\vec{k}_{1}\cdot(\vec{x}-\vec{y})}P(k_{1},t_{1}',t_{2}')\end{eqnarray}
where the prefactor is guessed from the physical interpretation of the
stress energy tensor and is defined to be independent of spacetime
(these quantities should be expressible in terms of $\vec{M}$ and
$T_{*}(\vec{M})$). This guess of the prefactor will be seen to be
important below since it gives us an argument for the order of
magnitude of $k^{3}P$. Given the assumptions of the previous
paragraph, we define the dimensionless function
$F_{2}$: \begin{equation} k^{3}P(k,t_{1}',t_{2}')\equiv
  F_{2}(k,T_{*},\Delta
  t,t_{1}'-t_{*},t_{2}'-t_{*},\vec{M})\end{equation} where we have
used Eq.~(\ref{eq:delttdefhere}), Eq.~(\ref{eq:wherehubbleenters}),
and assumption number 1 in the previous paragraph. Note that the
beginning of PT time $t_{*}$ only appears explicitly in the
combination $t_{i}-t_{*}$. Given that $F_{2}$ is a dimensionless
function, we can make $F_{2}$ out of the dimensionless combinations
shown in Table \ref{tab:Dimensionless-parameters}. Now, for
EWPT occurring at a temperature of order of $10^{2}$ GeV,
LISA and BBO are sensitive to wave vectors at the time of the phase
transition in the range \begin{equation}
  \frac{k}{a_*} \in[10^{-13},10^{-7}]\left(\frac{g_{*}(t_{0})}{3.9}\right)^{-1/3}\left(\frac{g_{*}(t_{*})}{10^{2}}\right)^{1/3}\left(\frac{T_{*}}{10^{2}\mbox{
      GeV}}\right)\mbox{ GeV}\end{equation} (corresponding to
$10^{-4}$ to $10^{2}$ Hz range detector frequencies).  Since we expect
$F_{2}\sim\mathcal{O}(1)$ as noted above, for unsuppressed dependence
on these parameters, we must have $\mathcal{O}(1)$ parametric
possibilities for these dimensionless parameters on which $F_{2}$
depends. For example, for $M_{i} a_* \Delta t\sim10^{14}$ to produce an
$\mathcal{O}(1)$ number consistent with $F_{2}$, we could have a term
proportional to\begin{equation} (M_{i}a_* \Delta t)^{-1/14}\end{equation}
in which case any scaling associated with $\Delta t$ will lead to a
suppressed change in $F_{2}$. Another possibility is an extreme fine
tuning of parameters\begin{equation} 1/(10^{-14}M_{i}a_* \Delta
t).\end{equation} This latter possibility cannot be excluded based on
dimensional analysis of the form here, and our scaling analysis can
break down if there are extremely large or small parameters. The
extreme level of fine tuning must be at least at the level of 1 part
in $10^{10}$ even when we allow for dynamically generated
dimensionless numbers of order $10^{4}$.  Finally, we can divide those
quantities in the table that are large with each other to give an
$\mathcal{O}(1)$ number.  Hence, from the table, barring an unlikely
dynamically generated fine tuning that we discussed, we conclude that
over the frequency range of interest for gravitational wave
detectors\begin{equation} F_{2}\approx F_{3}(k\Delta
t,\frac{t_{i}'-t_{*}}{\Delta
  t},\frac{\vec{M}}{T_{*}(\vec{M})}).\end{equation} Since the last
factor $\vec{M}/T_{*}(\vec{M})$ is independent of $H_{*}$ we arrive at
our robust approximation that the dimensionless spectrum $k^{3}P(k)$
only depends on the form given in Eq.~(\ref{eq:powerspec-scaling}).

\begin{table}
\begin{centering}
\begin{tabular}{|c|c||c|c|}
\hline 
quantity & range & quantity & range\tabularnewline
\hline
\hline 
$\frac{k/a_*}{T_{*}}$ & $[10^{-15},10^{-9}]$ & $T_{*}(t_{i}'-t_{*})a_*$ & $\mathcal{O}(10^{14})$\tabularnewline
\hline 
$k\Delta t$ & $[10^{-1},10^{3}]$ & $\frac{T_{*}}{M_{i}}$ & $\mathcal{O}(1)$\tabularnewline
\hline 
$k(t_{i}'-t_{*})$ & $\sim k\Delta t$ & $\frac{\Delta t}{t_{i}-t_{*}}$ & $\mathcal{O}(1)$\tabularnewline
\hline 
$\frac{k/a_*}{M_{i}}$ & $[10^{-15},10^{-9}]$ & $M_{i}a_*\Delta t$ & $\mathcal{O}(10^{14})$\tabularnewline
\hline 
$T_{*}a_*\Delta t $ & $\mathcal{O}(10^{14})$ & $M_{i} (t_{i}-t_{*})a_*$ & $\mathcal{O}(10^{14})$\tabularnewline
\hline
\end{tabular}
\par\end{centering}

\caption{\label{tab:Dimensionless-parameters}Dimensionless parameters with
$a_* \Delta t\sim10^{-2}/H_{*}\sim10^{12}$GeV$^{-1}$ and $T_{*}\sim M_{i}\sim100$
GeV. Since we expect $F_{2}\sim\mathcal{O}(1)$, for unsuppressed
dependence on these parameters, we must have $\mathcal{O}(1)$ parametric
possibilities for these dimensionless parameters on which $F_{2}$
depends if there is no fine tuned dimensionless parameters in the
theory.}

\end{table}

Next, we consider another possible scale, the turbulence's microscale
$\lambda$ , which is related to the largest scale as $\lambda=(Re)^{-3/4}L\sim10^{-10}L$.
In the short-lasting source's model \cite{Kamionkowski:1993fg,Caprini:2006jb},
this scale serves as a cut-off to the gravity wave spectrum. Since
this scale is $10^{10}$ higher than the peak scale, we can follow
the same reasoning about the short distance scale physics to show
that it would not be significant for the GW detector LISA or BBO.
In the recent work \cite{Caprini:2009yp} where long-lasting source
is considered, the microscale can affect the duration of the turbulence's
free decay. This duration of turbulence may appear as an independent
time-scale, but as we show in Appendix B, the decay duration is sufficiently
long that the turbulence has depleted most of its energy towards the
end of this duration, and the exact ending time, to an excellent approximation,
does not enter the GW spectrum. 

Another issue that we did not address are classical scaling violations
coming from quantum radiative corrections.  If the $\xi$ scaling is
many orders of magnitude, the anomalous dimension effect may give
(depending on what renormalization prescription is chosen) an
$\mathcal{O}(1)$ correction to the results presented here.  However,
in that case, GWs from first order PTs are much less likely to be
measurable and therefore are unlikely to be of practical interest.

Before we conclude the caveat section, we would like to comment on
the velocity dependence of the GW spectrum. In \cite{Kamionkowski:1993fg},
the GW spectrum's magnitude is given by Eq.\,(\ref{eq:KamionkowskiEq}).
The GW spectrum depends on the velocity $v_{w}$ implicitly through
$\alpha$ as for example in \begin{equation}
\kappa^{2}\left(\frac{\alpha}{1+\alpha}\right)^{2}\left(\frac{v_{w}^{3}}{0.24+v_{w}^{3}}\right), \end{equation}
since $v_{w}$ and $\kappa$ are functions of $\alpha$ as in
Eq.(\ref{eq:vw_alpha}) and (\ref{eq:kappa_alpha}). 
In the weak detonation limit of $\alpha\rightarrow0$,
we have $v_{w}\rightarrow\sqrt{1/3}$ and $\kappa\rightarrow\frac{4}{27}\sqrt{\frac{2}{3}\alpha}$
, causing the prefactor to scale as $\alpha^{3}$. In Caprini et
al. \cite{Caprini:2007xq}, the velocity dependence goes as \begin{equation}
v_{f}^{4}\frac{(1-s^{3})^{2}}{(1-(sv_{f})^{2})^{4}}\end{equation}
which in the weak detonation limit scales as $\alpha^{5}$ (since
$v_{f}\rightarrow\sqrt{3\alpha/2},s\rightarrow1-\sqrt{2\alpha})$.
For other region of parameters, as is shown in Fig.(13b) of \cite{Caprini:2007xq},
the two approaches can give peak amplitudes that differ by one order
of magnitude.  It is clear that the velocity dependence of the GW
spectrum is both uncertain and can be of high polynomial power, making
the current numerical uncertainty in the bubble wall velocity (see
e.g. \cite{Moore:2000wx}) a significant source of overall GW spectrum
uncertainty relevant for assessing the measurability of the GW.  It is
comforting however to know that our scaling rule is largely
independent of this uncertainty.

\section{\label{sec:summary}Summary}
In this paper, we have presented an analytic transformation rule
Eq.~(\ref{eq:spectrum shift}) that is useful for understanding how the
gravitational wave spectrum generated through an electroweak scale
first order phase transition at a fixed temperature $T_*$ would change
if the expansion rate of the universe during the phase transition were
different from that inferred from the assumption of pure radiation
domination.  We have explored the remarkable robustness of the scaling
relationship with respect to many computational uncertainties in the
gravity wave spectrum.

We apply this transformation rule to the example of a universe having a
quintessential kination dominated phase and find as expected a strong
sensitivity to the single phenomenological parameter controlling this
scenario.  In principle, this scaling relationship together with dark
matter properties measured by colliders can be used to overconstrain
this single phenomenological parameter.  Unfortunately, if the current
technology of gravity wave computations is correct, then we find that
any measurement of the gravity wave spectrum at the level of the projected BBO
sensitivity can rule out any appreciable boost factors relevant for
reconciliations between various sets of data such as that between
colliders and cosmology and/or indirect detection (such as that
relevant \cite{Chung:2007vz} for PAMELA data \cite{Adriani:2008zr}).

Nonetheless, using the results of this work, any future gravity wave
detection experiments measuring phase transition induced gravity waves
can understand their measurement's sensitivity to the expansion rate
of the universe.  It would indeed be exciting to have an observational
anchor on the expansion rate of the universe when the universe is as
hot as $100$ GeV, just as isotope abundance measurements allow us to
have an observational anchor on the expansion rate at a temperature of
$1$ MeV in the context of big bang nucleosynthesis.
\section*{Acknowledgments}

We thank R.~Durrer, L.~Everett, T.~Konstandin, A.~Kusenko, and G.~Servant for
useful conversations.  DJHC thanks Galileo Galiei Institute (GGI) where part of
this work was completed. (Preliminary results of this work was
presented at GGI 1/21/09.) We also thank J.~Schmitthenner for
initial collaboration on this project. This project is supported by
DE-FG02-95ER40896.

\appendix
\section{\label{sec:Bubbles-Filling-Space}Bubbles Filling Space}

Here we review a well known argument \cite{Guth:1980zk} about how
first order PT bubbles fill space.  For this section, we will use the
metric parameterization $ds^2=dt^2 -a^2(t)|d\vec{x}|^2$.  Let
$\gamma(t)$ denote the probability per volume per time of bubble
fomation. Assuming that the bubble wall is not accelerating in the
locally inertial frame (which is what is typically done in the
literature when the bubble wall velocity $v_{w}$ is taken to be a
constant), we have \begin{equation}
  \frac{d\vec{x}}{d\tau}\propto\frac{1}{a}.\end{equation} Hence, we
have\begin{equation}
\frac{1}{\sqrt{1-a^{2}(\frac{dr}{dt})^{2}}}\frac{dr}{dt}=\frac{K}{a}\end{equation}
where the constant $K$ is to be determined by a boundary condition.
Setting the boundary condition\begin{equation}
a_{i}\frac{dr}{dt}|_{t_{i}}=v_{w},\end{equation} we
find\begin{equation} K=\frac{v_{w}}{\sqrt{1-v_{w}^{2}}}\end{equation}
and\begin{equation} \frac{dr}{dt}=\frac{v_{w}}{a}\end{equation} Hence,
a bubble nucleated at time $t_{i}$ fills a comoving
volume\begin{equation}
V_{3}(t_{i},t)=\frac{4\pi}{3}\left[v_{w}\int_{t_{i}}^{t}\frac{dt'}{a(t')}\right]^{3}.\end{equation}

Next, we compute $P$, the probability that a point in comoving space
is in false vacuum. The probability that at time $t+dt$ the vacuum
at a point is still in false vacuum given that it is in false vacuum
at time $t$ is \begin{equation}
P(t+dt)=P(t)[1-P_{c}]\end{equation}
 where $P_{c}$ is the probability of nucleating a bubble within the
past causal cone surface volume of thickness $dt$ with the causal
signal propagation speed given by $v_{w}$ since it is the bubble
wall that needs to reach the given point in consideration. Since the
relevant surface volume of the causal cone can be easily computed
to be\begin{equation}
a^{3}V_{3}(t_{i},t)dt,\end{equation}
 we can multiply this by $\gamma$ to find $P_{c}$ to arrive at\begin{equation}
\frac{dP}{dt}=-Pa^{3}V_{3}(t_{i},t)\gamma(t).\end{equation}
 Solving for $P$, we find\begin{equation}
P(t)=P(t_{i})\exp\left(-\int_{t_{i}}^{t}dt'\gamma(t')V_{3}(t_{i},t')a^{3}(t')\right).\end{equation}

Using Eq.~(\ref{eq:nucleationrate}), we find\begin{equation}
P(t)=P(t_{i})\exp\left(-C_{1}\int_{t_{i}}^{t}dt'\exp\left[\left(-S_{*}^{(3)}+\frac{(t'-t_{*})H_{*}}{1+\frac{1}{3}\frac{d\ln g_{*S}}{d\ln T}}\frac{dS^{(3)}}{d\ln T}|_{t_{*}}\right)/T(t')\right]T^{4}(t')V_{3}(t_{i},t')a^{3}(t')\right).\end{equation}
 When $t<t_{c}$ where\begin{equation}
t_{c}\sim t_{*}+\left[\frac{H_{*}}{1+\frac{1}{3}\frac{d\ln g_{*S}}{d\ln T}}\frac{d\ln S^{(3)}}{d\ln T}|_{t_{*}}\right]^{-1},\end{equation}
 $S_{*}^{(3)}/T$ is large such that the exponential suppression in
the integrand makes the integral in front of $C_{1}$ negligible.
After that time scale, probability of not being in the false vacuum
is $\mathcal{O}(1)$, and the PT is assumed to be completed.
Hence, we can conclude that the duration of the PT
scales as\begin{equation}
\Delta t_{\mbox{proper}} \equiv
t_{c}-t_{*}\propto\frac{1}{H_{*}}.\label{eq:durationofPT}
\end{equation} 
In terms of conformal time, when $H_* \Delta t_{\mbox{proper}} \ll 1$, we have
\begin{equation}
\Delta t_{\mbox{conformal}} \approx \Delta
t_{\mbox{proper}}/a(t_*)\propto \frac{1}{H_*}.
\end{equation}
The variable $\Delta t_{\mbox{conformal}}$ corresponds to the variable
$\Delta t$ in Eq.~(\ref{eq:proportionality}).

\section{The decay duration of turbulence}

In the GW spectrum formula Eq.\,(\ref{eq:CDSturb}) in reference
\cite{Caprini:2009yp}, the free decay part of turbulence gives a
contribution that proportional to the integral \begin{equation}
  \int_{1}^{y_{fin}}dy\frac{y^{-7\gamma_*}}{y+\frac{t_{in}}{\tau_{L}}}I_{s}(K_{*},y,y)\int_{y}^{y_{top}}\frac{dz}{z+\frac{t_{in}}{\tau_{L}}}\cos\left(\frac{\pi
    K_{*}}{v_{L}}(z-y)\right)\end{equation} Here, $y$ and $z$ are
dimensionless time variables: e.g. $y=(t-t_{in})/\tau_{L}$ where $t_{in}$
is the beginning time of the stirring phase and $\tau_{L}$ is the
largest eddy turn over time.  We now show that the ending time of
turbulence as a function of the scale length $y_{fin}(k)$ is very
large for the peak position $(y_{fin}(k_{peak})\sim10^{4})$ and
therefore the exact ending time is irrelevant to GW spectrum at the
peak position. We shall consider the behavior of the integrand as
$y\rightarrow y_{fin}$ , and take $\gamma_*=2/7$ for
concreteness:\begin{equation} 
  \int_{1}^{y_{fin}}dy\frac{y^{-7\gamma}}{y+\frac{t_{in}}{\tau_{L}}}I_{s}(K_{*},y,y)\int_{y}^{y_{top}}\frac{dz}{z+\frac{t_{in}}{\tau_{L}}}\cos\left(\frac{\pi
    K_{*}}{v_{L}}(z-y)\right)
 \sim \int_{1}^{\infty}dy\, y^{-5}-O(y_{fin}^{-4})\end{equation} where we
have used the following estimation \begin{align*} I_{s}(K_{*},y,y) &
  \longrightarrow\begin{cases} y^{-1.05} & \mbox{MHD
    turbulence}\\ y^{-1} & \mbox{fluid turbulence}\end{cases}\sim
  y^{-1}\\ y_{top}-y & =\min[y_{fin}-y,\frac{x_{c}v_{L}}{\pi
      K_{*}}]\sim v_{L}/K_{*}\sim O(1)\end{align*} Therefore the GW's
relative dependence on $y_{fin}$ is as weak as $y_{fin}^{-4}$ . For
the ending time of turbulence at the stirring scale
$K_{*}=kL_{*}\sim1$ , one can use Eq.~(72) of \cite{Caprini:2009yp}
\begin{equation}
y_{fin}\sim\frac{t_{fin}(k)}{\tau_{L}}\approxeq\left(\frac{2\times10^{14}}{K_{*}}\right)^{\frac{28}{101}}\mbox{, for }K_{*}>0.07\end{equation}
which shows that $y_{fin}(K_{*}=1)\sim10^{4}$. If one take
$y_{fin}\rightarrow\infty$ in the above integral, the error introduced
is on the order of $O(10^{-16})$, very small compared to the integral
itself which is at least $O(1)$.

\section{\label{sec:formalmap}A Formal Map between $V(q)$ and $n$}

Here we give a formal map between the quintessence potential $V(q)$
and the quintessence energy density dilution behavior $a^{-n}$. 

The equation of motion of the quintessence can be written using the
scale factor $a$ as a time variable assuming that the only other
component during the era of interest is radiation which dilutes as
$a^{-4}$:\begin{equation}
\frac{1}{a^{3}}\frac{\sqrt{2}\sqrt{V(q)/\rho_{R0}+(\frac{a_{0}}{a})^{4}}}{\sqrt{6-\frac{1}{M_{p}^{2}}(\frac{dq}{d\ln a})^{2}}}\frac{d}{d\ln a}\left(\frac{\sqrt{2}\sqrt{V(q)/\rho_{R0}+(\frac{a_{0}}{a})^{4}}}{\sqrt{6-\frac{1}{M_{p}^{2}}(\frac{dq}{d\ln a})^{2}}}\frac{d(q/M_{p})}{d\ln a}a^{3}\right)+\frac{M_{p}V'(q)}{\rho_{R0}}=0.\end{equation}
The solution to this equation defines a functional\begin{equation}
q_{s}[a,V(q)].\end{equation}
Note that the equation has been normalized to be dimensionless: $V$
is measured in units of initial radiation energy density $\rho_{R0}$
and $q$ is measured in units of $M_{p}$. Next, one can solve $
q=q_{s}[a,V(q)]$ for $a[q,V(q)]$.
This can be put into the energy density scaling equation for the quintessence
energy density, yielding the equation\begin{equation}
\frac{\rho_{q0}}{\rho_{R0}}(\frac{a_{0}}{a[q,V(q)]})^{n}=\frac{1}{2}\left[\frac{\sqrt{2}\sqrt{V(q)/\rho_{R0}+(\frac{a_{0}}{a[q,V(q)]})^{4}}}{\sqrt{6-(\frac{dq_{s}[a,V(q)]/M_{p}}{d\ln a})^{2}}}\right]^{2}(\frac{dq_{s}[a,V(q)]/M_{p}}{d\ln a})_{a=a[q,V(q)]}^{2}+\frac{V(q)}{\rho_{R0}}\end{equation}
which can in principle be solved for $V(q)$.

\section{\label{sec:TdepofS3} An Estimate of Temperature Dependence of
  $S^{(3)}$ } In this section, we will give a semi-quantitative
argument of which parameter choices of effective potential governing
the PT will lead to an enhanced gravity wave amplitude.
Consider the high temperature expansion of the effective potential for
a single real field in the form \begin{equation}
  V(\phi)=\frac{1}{2}(\mu^{2}+cT^{2})\phi^{2}-E\phi^{3}+\frac{\lambda}{4}\phi^{4}\end{equation}
where $\mu^{2}<0$ and $c$ is a thermal correction dependent parameter.
Equation of motion yields\begin{equation}
0=\frac{1}{L^{2}}\phi+(\mu^{2}+cT^{2})\phi-3E\phi^{2}+\lambda\phi^{3}.\end{equation}
where we have estimated $\nabla^{2}\phi\sim\frac{1}{L^{2}}\phi$.  In
solving for $L$, we need a characteristic value for $\phi$ which we
will call $\phi_{c}$. We can set this characteristic value to be
between the local maximum $\phi_{u}$ and the minimum $\phi_{*}$ (not
the one at the origin) of the effective potential. (Recall that the
cubic term $-E\phi^{3}$ is responsible for there being a bump in the
potential giving rise to a local maximum.) Explicitly solving
$V'(\phi)=0$, we find\begin{equation}
\phi_{u}(T)=\frac{3E}{2\lambda}\left(1-\sqrt{1-\frac{4\lambda}{9E^{2}}(\mu^{2}+cT^{2})}\right)\end{equation}
and\begin{equation}
\phi_{*}(T)=\frac{3E}{2\lambda}\left(1+\sqrt{1-\frac{4\lambda}{9E^{2}}(\mu^{2}+cT^{2})}\right).\end{equation}
Hence, if we make a somewhat arbitrary but reasonable definition for
the characteristic value to be \begin{equation}
  \phi_{c}\equiv\phi_{u}(T)+\frac{\phi_{*}(T)-\phi_{u}(T)}{2},\end{equation}
we find\begin{equation} \phi_{c}=\frac{3E}{2\lambda}\end{equation}
independently of the temperature except through $E$ and $\lambda$
which we assume to be dominated by the non-thermal contribution ($E$
coefficients that rely on thermal corrections do not yield strong
phase transitions typically anyway). This yields the length scale
associated with the bubble action to be \begin{equation}
  L=\frac{1}{\sqrt{\frac{9E^{2}}{4\lambda}+|\mu^{2}|-cT^{2}}}\end{equation}
where we have displayed our assumption of $\mu^{2}<0$ manifestly.
Hence, we have\begin{eqnarray} S^{(3)} & \sim &
L^{3}\left((\frac{\phi_{c}}{L})^{2}+V(\phi_{c})\right)\\ & = &
\frac{\phi_{c}^{2}}{\sqrt{\frac{9E^{2}}{4\lambda}+|\mu^{2}|-cT^{2}}}\\ &
&
+\frac{\frac{1}{2}(\mu^{2}+cT^{2})\phi_{c}^{2}-E\phi_{c}^{3}+\frac{\lambda}{4}\phi_{c}^{4}}{\left(\frac{9E^{2}}{4\lambda}+|\mu^{2}|-cT^{2}\right)^{3/2}}.\end{eqnarray}
Computing the temperature derivative at the critical temperature of
\begin{equation}
T=T_{c}=\frac{\sqrt{2E^{2}+\lambda|\mu^{2}|}}{\sqrt{c\lambda}}\end{equation}
(which is obtained by setting $V(\phi_{*})=0$ and solving for $T)$ we
find\begin{eqnarray} \frac{dS^{(3)}}{dT}|_{T=T_{c}} & \sim &
50\frac{\sqrt{c}}{\lambda}\sqrt{2+\lambda|\mu^{2}|/E^{2}}.\end{eqnarray}
Hence, we conclude\begin{eqnarray} \frac{\beta}{H_{*}} & \sim &
\frac{50}{1+\frac{1}{3}\frac{d\ln g_{*S}}{d\ln
    T}}\frac{\sqrt{c}}{\lambda}\sqrt{2+\lambda|\mu^{2}|/E^{2}}.\end{eqnarray}
To make this order unity (appropriate for an enhanced gravity wave
amplitude), we can consider $\sqrt{c}\ll1$. Assuming
$\lambda\sim\mathcal{O}(1)$, we find\begin{equation}
c\lesssim10^{-2}.\end{equation} This leads to a PT
temperature of\begin{equation}
T_{c}\gtrsim\mbox{TeV}\left(\frac{|\mu|}{100\mbox{
    GeV}}\right)\frac{\sqrt{2E^{2}/|\mu|^{2}+\lambda}}{\sqrt{\lambda}}\end{equation}
corresponding to a high temperature PT of a weakly coupled scalars.

\end{document}